\documentclass[referee]{raa}            

\usepackage{graphicx,times}             
\usepackage{natbib}
\usepackage{amssymb,amsmath}
\bibpunct{(}{)}{;}{a}{}{,}

\usepackage[a4paper=true,dvipdfm=true,pagebackref=true]{hyperref}
\hypersetup{colorlinks = true, linkcolor = green, anchorcolor = red, citecolor = blue, filecolor = red, pagecolor = red, urlcolor = red}

\begin{document}

   \title{A magnetic confinement nuclear fusion mechanism for solar flares
}

   \volnopage{Vol.0 (20xx) No.0, 000--000}      
   \setcounter{page}{1}          

   \author{Ying-Zhi Zhang
   }

   \institute{NO.1 Nanertiao, Zhongguancun, Haidian district, Beijing, 100190 China; {\it yzzhangmail@sohu.com}\\
\vs\no
   {\small Received~~20xx month day; accepted~~20xx~~month day}}

\abstract{ We propose a magnetic confinement nuclear fusion mechanism for the
  evolution of a solar flare in solar atmosphere. The mechanism agree with two
  observed characteristics of explosive flares and coronal mass ejections (CMEs) that
  have proved to be very difficult to explain with previous mechanisms: the huge
  enrichments of $^{3}He$ and the high energy gamma ray radiation. The twisted magnetic
  flux rope is a typical structure during the solar flares, which is closely related to
  the solar active region that magnetic fields have almost complete control over the
  plasma. Consequently, the plasma inside the flux rope is heated to more than
  1.0$\times10^{7}$ K by adiabatic compression process, and then the thermonuclear fusion
  can take place in the flux rope accompanied with high energy gamma rays. We utilize
  the time-dependent ideal 2.5-dimensional magnetohydrodynamic (MHD) simulation to
  demonstrate the physical mechanism for producing flares, which reveals three stages
  of flare development with process of magnetic energy conversion and intense release
  during the solar flares and CMEs in solar atmosphere. Furthermore, we discuss the
  relationship between magnetic reconnection and solar eruptions.
}

   \authorrunning{Y.-Z. Zhang }            
   \titlerunning{A magnetic confinement nuclear fusion mechanism for solar flares}  

   \maketitle

%
%
\section{Introduction}           
\label{sect:intro}
   Solar flares are an intense explosion that usually occur in complex magnetic field
configuration and has highly twisted magnetic flux rope structure, which are strongly
associated with active region filament eruptions and coronal mass ejections (CMEs)
(Jing et al. 2004, Toriumi et al. 2017). Both ground and space telescope observations
show that the solar flares exhibit a variety of phenomena, which they cause electromagnetic
radiations ranging from kilometric radio waves to tens of MeV gamma rays (Bastian et
al. 1998, Gruber et al. 2011, Ackermann et al. 2014, Reid \& Ratcliffe 2014, Benz 2017),
and they produce energetic particles escaping into interplanetary space (Bai \& Sturrock 1989,
Hudson \& Ryan 1995, Lin 2011, Reames 2013, Trottet et al. 2015, Dierckxsens et al. 2015,
Cliver 2016). Specifically, a considerable puzzle is the production mechanism of the
dramatically enrichments (up to a factor of ten thousand) of $^{3}He$ that is associated
with gamma rays burst during the solar flares (Schaeffer \& Z$\ddot{a}$hringer 1962, Hsieh
\& Simpson 1970, Kocharov \& Kocharov 1984, Mason et al. 2004, Mason 2007, Nitta et al. 2015).

   Following observational results, MHD model of the solar flares has experienced three
relatively clear development stages. In the first phase, the studies have not been able to
explain the time scale of a flare, which the dominant role of magnetic field in flare process
has been confirmed (Hale 1908), and several important features of flare (neutral point,
current sheet and magnetic reconnection, etc.) have been proposed (Giovanelli 1946,
Parker 1957, Sweet 1958). In the second phase, due to the establishment of the standard
flare model (Carmichael 1964, Sturrock 1966, Hirayama 1974, Kopp \& Pneuman 1976), the
energy storage model is the widely accepted model of a solar eruption. In the third
phase, the magnetic flux rope model has been investigated extensively (Low 1996, Priest \&
Forbes 2002, Longcope 2005, Shibata \& Magara 2011), which includes emerging magnetic
flux rope model (van Tend \& Kuperus 1978, Aly 1984, Forbes \& Isenberg 1991, Isenberg
et al. 1993, Forbes \& Priest 1995, Gibson \& Low 1998, Hu \& Liu 2000, Chen \& Shibata
2000, Lin et al. 2001, Fan 2001, Fan \& Gibson 2003, Zhang et al. 2005, Zhang \& Wang 2007)
and shearing magnetic flux rope model (van Ballegooijen \& Martens 1989, Mikic \& Linker 1994,
Antiochos 1998, Antiochos et al. 1999, Titov \& D$\acute{e}$moulin 1999, Amari et al. 2000,
$T\ddot{o}r\ddot{o}k$ et al. 2004).

   In fact, the physical process of making the abundant $^{3}He$ accompanied by the high
energy gamma rays usually implies the thermonuclear reaction to occur in the solar flares.
Furthermore, it is critical to producing enough high temperature plasma for the thermonuclear
fusion. Therefore, we propose a magnetic confinement nuclear fusion mechanism for dealing
with energy conversion and intense release during the solar flares in solar atmosphere.
The magnetic energy may be converted into plasma thermal energy stored in the magnetic flux
rope by adiabatic compression process, and the tremendous enrichment of $^{3}He$ and high
energy gamma rays are produced by nuclear fusion during the magnetic confinement process.
The purpose of this article is to discuss the details of the new mechanism for flare
generation in the solar atmosphere.

\section{Theoretical Basis}
\label{sect:Theo}

\subsection{P-p chain reaction in the sun}

Hydrogen is the most abundant element in the main-sequence stars, which it can undergo
thermonuclear reaction at temperature greater than 7$\times10^{6}$K, and p-p chain reaction
is the dominant fusion reaction pattern at temperature lower than 1.5$\times10^{7}$ K in
the sun. It is now well established that the reactions making up the overwhelming bulk of
the p-p chain are the following:
$$
^{1}H + ^{1}H \longrightarrow ^{2}H + e^{+} + \nu_{e} + 1.44MeV,
\eqno (1)
$$
$$
^{2}H + ^{1}H \longrightarrow ^{3}He + \gamma + 5.49MeV,
\eqno (2)
$$
$$
^{3}He + ^{3}He \longrightarrow ^{4}He + ^{1}H + ^{1}H + 12.86MeV.
\eqno (3)
$$
The rate of the reaction chain as a whole and hence the rate of energy production are
controlled by the first and third reactions which take place very slowly. By contrast,
the second reaction is extremely fast, effectively converting $^{2}H$ to $^{3}He$ only
in a few seconds. At temperature T in the neighborhood of a temperature $T_{r}$ (in
1$\times10^{6}$K), the rate of energy production $\epsilon$ can be expressed in the
form (Salpeter 1952):
$$
\epsilon = \epsilon_{0} \frac{\rho x^{2}_{H}}{100} (\frac{T}{T_{r}})^{n} erg\cdot g^{-1}\cdot s^{-1},
\eqno (4)
$$
where $\rho$ is the density, and $x_{H}$ is the concentrations (by mass) for
hydrogen. For various temperature T, the exponent of the variation with temperature n
and the rate of energy production $\epsilon_{0}$ are given in Table 1 by Salpeter
(1952). If the chain goes only to $^{3}He$, as might be the case in limited times at low
temperature, then the rate of energy production in $3\times^{1}H\longrightarrow^{3}He$
is $\epsilon_{pp}^{'}$ = 0.509$x_{H}^{-2}$$\epsilon_{pp}$ (Burbidge et al. 1957).

Although we know that nuclear fusion usually occurs in the solar core, we can infer
that the p-p chain reaction tend to take place during the flare process in the solar
magnetic activity, for which the key ingredient in the solar flares is the anomalous
overabundances of $^{3}He$ accompanied by MeV gamma rays. The astronomical evidence
suggests that nuclear reactions can take place in regions (such as stellar magnetic
activity) on the stellar surface, for which the regions develop sufficient magnetic
energy to accelerate particles (Fowler et al. 1955). We conclude that the
plasma can be heated to much more than 7.0$\times10^{6}$K by a magnetic confinement
mechanism in the solar active region, and then $^{3}He$ is the end product of
nuclear reaction since solar flares have lasted only a few hours in the solar
atmosphere. Under astrophysical conditions, we assume that the equation used to
calculate the rate of energy production of nuclear reaction at the interior of the
sun is also applicable to calculate that at the solar surface. Then for temperature $T_{f}$
and density $\rho_{f}$ in the core of solar flares, the rate of energy
production $\epsilon_{f}$ in the flare process can be expressed in the form:
$$
\epsilon_{f} = 2.545\times10^{-25}\rho_{f}T_{f}^{4} erg\cdot g^{-1}\cdot s^{-1}\cdot K^{-4}.
\eqno (5)
$$

\subsection{Magnetic confinement in the solar atmosphere}

Due to the higher conductivity in the solar atmosphere, ideal MHD method is suitable for
studying the evolution of the solar flare events. Furthermore, magnetic flux emergence
from the solar photosphere into the corona is the driver of a variety of phenomena and
the inclusion of different physical effects (such as magnetic flux rope formation, current
sheet, and magnetic reconnection) associated with solar activity (Cheung \& Isobe 2014).
We consider that the magnetic flux rope suspended in the corona with a transverse current
sheet above and a vertical current sheet below may be the magnetic confinement structure
in solar atmosphere.

The solar flares start with a magnetic flux rope formatted by magnetic emergence. The plasma
inside the twisted magnetic flux rope is adiabatically heated by the helical magnetic field.
We conclude that some of the magnetic energy is converted into plasma thermal energy by
adiabatic compression and the remainder of the magnetic energy has strengthened the magnetic
flux rope structure during the process of the magnetic confinement. For completely ionized
plasma in the corona, suppose ions and electrons inside the flux rope are at the same
temperature (named plasma temperature). While the temperature of the pinched plasma is
greater than 7$\times10^{6}$K, the thermonuclear fusion can take place in the magnetic flux rope.

Finally, the magnetic confinement structure is destroyed by magnetic reconnection. While
current density of vertical current sheet reaches its peak value, the magnetic reconnection
may occur in the current sheet associated with nuclear explosion, which is similar to the
phenomenon of short circuit in conductors. The magnetic reconnection is a key ingredient of
many astrophysical phenomena, which the release of stored magnetic energy is facilitated by
the ideal MHD process (Pontin 2012). Furthermore, free magnetic energy of the flux
rope system is converted into plasma kinetic energy (CMEs) triggered by magnetic reconnection.
Specifically, because of thermonuclear reactions, particles get kinetic energy
not by some kinds of acceleration mechanism, but by high temperature.

\section{Numerical Model}
\label{sect:num-model}

We utilize time-dependent ideal 2.5-dimensional MHD simulation of the two-stage catastrophic
magnetic flux rope model to illustrate the evolution of the solar flares (Zhang 2013, 2015).
In spherical coordinates (r,$\theta,\varphi$), a magnetic flux function $\psi(t,r,\theta)$
was presented, and it has to do with magnetic field by

$$
{\mathbf B} = \bigtriangledown\times \left (
\frac{\psi}{r\sin\theta}\hat{\varphi} \right ) + {\mathbf B}_\varphi,
\ \ \ \ {\mathbf B}_\varphi = B_\varphi \hat{\varphi} ,
\eqno (6)
$$
where $B_\varphi$ is the azimuthal component of the magnetic field. The ideal MHD equations
are the following form:
$$
\frac{\partial\rho}{\partial t} + \bigtriangledown\cdot(\rho {\mathbf v})=0,
\eqno (7)
$$
$$
\frac{\partial{\mathbf v}}{\partial t} + {\mathbf
v}\cdot\bigtriangledown {\mathbf v} +
\frac{1}{\rho}\bigtriangledown p + \frac{1}{\mu\rho}
[L\psi\bigtriangledown\psi + {\mathbf
B}_\varphi\times(\bigtriangledown \times{\mathbf B}_\varphi )]
$$
$$
+\frac{1}{\mu\rho r\sin\theta}\bigtriangledown\psi
\cdot(\bigtriangledown\times{\mathbf B}_\varphi )\hat{\varphi} +
\frac{GM_{\odot}}{r^2}\hat{r}=0,
\eqno (8)
$$
$$
\frac{\partial\psi}{\partial t}+ {\mathbf
v}\cdot\bigtriangledown\psi = 0,
\eqno (9)
$$
$$
\frac{\partial B_{\varphi}}{\partial t} +
r\sin\theta\bigtriangledown \cdot \left ( \frac{B_\varphi{\mathbf
v}}{r\sin\theta} \right ) + \left [ \bigtriangledown \psi\times
\bigtriangledown \left ( \frac{v_\varphi}{r\sin\theta} \right )
\right ]_\varphi = 0,
\eqno (10)
$$
$$
\frac{\partial T}{\partial t}+{\mathbf v}\cdot\bigtriangledown T +
(\gamma-1)T\bigtriangledown\cdot{\mathbf v} = 0,
\eqno (11)
$$
where
$$
L\psi\equiv\frac{1}{r^{2}\sin^{2}\theta}
\left ( \frac{\partial^{2}\psi}{\partial
r^{2}}+\frac{1}{r^{2}}\frac{\partial^{2}\psi}{\partial\theta^{2}}-
\frac{\cot\theta}{r^{2}}\frac{\partial\psi}{\partial\theta} \right ) ,
\eqno (12)
$$
For an ideal adiabatic compression process, the polytropic index is $\gamma = 2.3$.
The solution domain is $1 \leq r \leq 30$, $0\leq\theta\leq \pi/2$. It is discretized
into $130\times 90$ grid points. The above MHD equations are solved with the
multi-step implicit scheme developed by Hu (1989).

In this study, we choose the essential physical parameters of bottom of the corona
above the active region, which the temperature $T_0=3\times 10^6$ K and the density
$\rho_0 = 1.67\times 10^{-11}$kg$\cdot$m$^{-3}$. In the following, the solar radius
$R_s$ is the unit of length, and the ration of gas pressure to magnetic
pressure is $\beta$ = 0.01, so that the unit of $\psi$ is
$\psi_0$ = $(2\mu \rho_0 RT_0 R_s^4 / \beta )^{1/2}$ = 6.97$\times 10^{15}$ Wb, and
the unit of magnetic field intensity is $B_0$ = $\psi_0/R_s^2$ = 1.44$\times 10^{-2}$ T.
Some other units of interest are $E_0$ = $B_0^2R_s^3/\mu$ = 5.563 $\times 10^{28}$ J
for magnetic energy, $v_A$ = $B_0/(\mu\rho_0)^{1/2}$ = 3150 km$\cdot$s$^{-1}$ for
velocity, $\tau_A$ = $R_s/v_A$ = 220 s for time, and $j_0 = B_0/(\mu R_s)$ =
1.65$\times 10^{-5}$A$\cdot$m$^{-2}$ for current density.

A quadrupole field is chosen for the initial background magnetic field (see Fig. 3a).
The related magnetic flux function normalized by $\psi_0$ is given by
$$
\psi(r,\theta)=\frac{\sin^{2}\theta}{r}+\frac{(3+5\cos2\theta)
\sin^{2}\theta}{2r^{3}} ,
\eqno (13)
$$
where $r$ is in the unit of $R_s$. During the evolution of the magnetic flux rope
system, two stage catastrophes occur in the solar corona (Zhang 2015). However,
the whole process has been without the plasma macroscopic instability, and the
magnetic reconnection occurs in the vertical current sheet while the current density
gets the maximum value.

The magnetic energy ($E$) of the force-free magnetic field is normalized by $4\pi E_0$,
and calculated by the following equation:
$$
E = \frac{1}{2} \int_1^{30}  dr \int_0^{\pi /2}
B^2 r^2\sin\theta d\theta + \frac{30^3}{2}
\int_0^{\pi /2} (B_r^2 - B_\theta^2 )_{r=30} \sin\theta d\theta ,
\eqno (14)
$$
where the first term on the right is the magnetic energy in the computational domain
(1 $\leq r \leq$ 30; $0 \leq \theta \leq \pi$; $0 \leq \varphi \leq 2\pi$) and the
second is outside the domain (see Hu 2004). The magnetic energy is $E_p$ = 1.476 for
the initial quadrupolar potential field, and $E_m$ = 1.662 for the corresponding partly
open field (see Zhang et al. 2005). Moreover, the magnetic energy inside the flux rope
($E_r$) is calculated by the representation similar to the first term on the right of
the equation (14), but with the integral domain limited to the interior of the flux
rope.

\section{Numerical Result}
\label{sect:num-result}

In our simulation, the magnetic flux rope is described with its toroidal magnetic
flux ($\Phi_{p}$) and poloidal flux ($\Phi_{\varphi}$), which are deterministic
throughout the magnetic flux emergence in the evolution of the magnetic flux rope
system. In present study, the initial condition is in an equilibrium state,
$\Phi_{p}$ = 0.42 and $\Phi_{\varphi}$ = 0.0544. Base on the result, we illustrate the
key characteristics of flare development, such as magnetic flux rope for the formation
of magnetic confinement structure, high temperature plasma for thermonuclear fusion
and magnetic reconnection associated with nuclear explosion in the corona. Furthermore,
the two stage catastrophes are the mechanism driving the eruption of magnetic flux
rope in the simulation.

\subsection{Formation of the magnetic confinement structure}

Two parameters are used to describe the geometrical characteristics of the flux rope
system in equilibrium. One is the height of the flux rope axis, $h_a$, and the other is
the length of the vertical current sheet, $h_c$. Figure 1 shows that there are two
stages of catastrophes during the evolution of the magnetic flux rope system in the
solar corona, and the flux rope has maintained a certain period of relative stability
between the two catastrophes. The magnetic flux rope has emerged from the photosphere
into the solar corona after 1$\tau_{A}$. Then the first catastrophe occurs in the flux
rope system at 4$\tau_{A}$, which the vertical current sheet appears below the flux rope.
At 39$\tau_{A}$, the second catastrophe occurs with the flux rope escaping.

\begin{figure}
  \centering
  \includegraphics[]{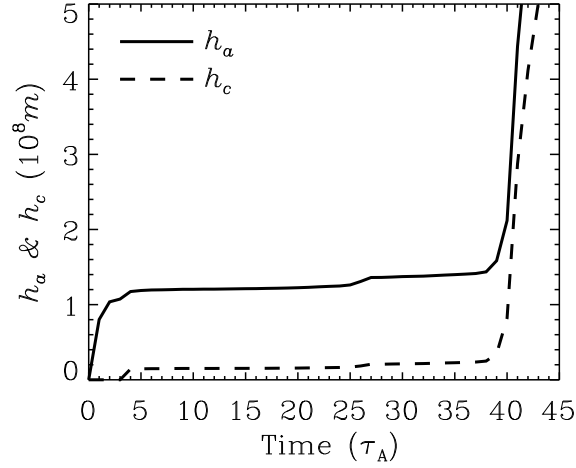}
  \caption{
  Height of the flux rope axis, $h_a$ vs. time (solid lines), and height of the
  vertical current sheet under the flux rope, $h_c$ vs. time (dashed lines).
  }\label{Fig1}
\end{figure}

As seen from Figure 2, the plasma temperature at the flux rope axis, $T_a$, and the plasma
density at the flux rope axis, $\rho_a$, are represented during the flare evolution in the
solar corona, respectively. Since the plasma is subjected to adiabatic compression by the
magnetic field, $T_a$ has been greater than 1$\times10^{7}$ K from 1$\tau_{A}$ to 41$\tau_{A}$
(Fig. 2a). At the early stage of magnetic flux rope eruption, a hot channel with a
temperature as high as 1$\times10^{7}$ K along the magnetic flux rope was found by SDO/AIA
observations (see Zhang et al. 2012). This numerical result may explain the observations.
While the flux rope just emerges into the solar corona at 1$\tau_{A}$, $T_a$
reaches its maximum value, 3.494$\times10^{7}$ K (see Fig. 3b). We conclude that the
plasma can be ignited and thermonuclear reaction takes place inside the flux rope because this
temperature (3.494$\times10^{7}$ K) is much higher than 7$\times10^{6}$K. After that, $T_a$
plummets to 1.857$\times10^{7}$ K just before the first catastrophe at 3$\tau_{A}$, and drops
from 1.950$\times10^{7}$ K to 1.067$\times10^{6}$K until the eruption at 41$\tau_{A}$. During
the evolution of the flux rope system, $\rho_a$ has remained at about
2.0$\times 10^{-11}$$kg\cdot m^{-3}$ except before the first catastrophe and after the
second catastrophe (Fig. 2b).

\begin{figure}
  \centering
  \includegraphics[]{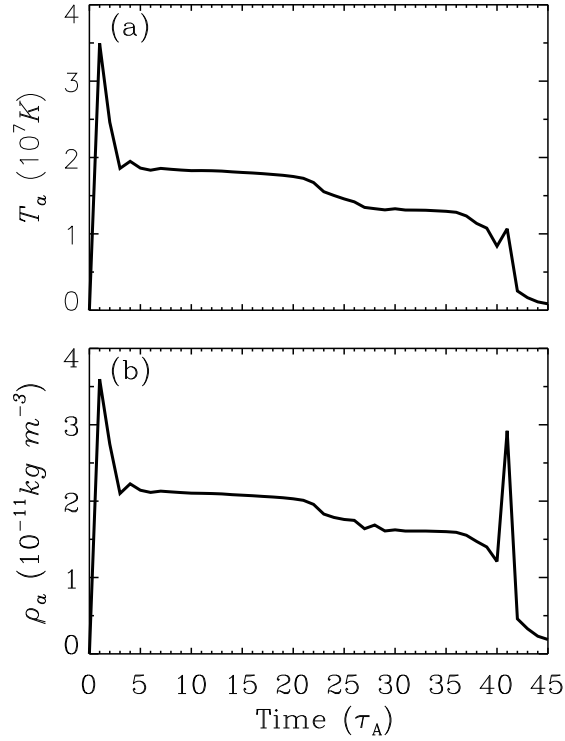}
  \caption{
  Plasma temperature at the flux rope axis, $T_a$ vs. time (a), and plasma
  density at the flux rope axis, $\rho_a$ vs. time (b).
  }\label{Fig2}
\end{figure}

Furthermore, after the first catastrophe occurs in the magnetic flux rope system, a vertical
current sheet is formed below the flux rope, and a curved transverse current sheet developed
from a neutral point is formed above the flux rope, which the high temperature region appears
near the two footpoints of the transverse current sheet (about 8$\times10^{6}$ K) and inside
the magnetic flux rope (greater than 1$\times10^{7}$ K), respectively (see Fig. 3c).
We further conclude that the magnetic confinement structure composed of a magnetic flux rope with two
current sheets has been formed in the corona after the first catastrophe, which the high
temperature plasma heated by an adiabatic compression process in the twisted magnetic flux
rope is involved in the thermonuclear fusion reaction.

\begin{figure}
  \centering
  \includegraphics[]{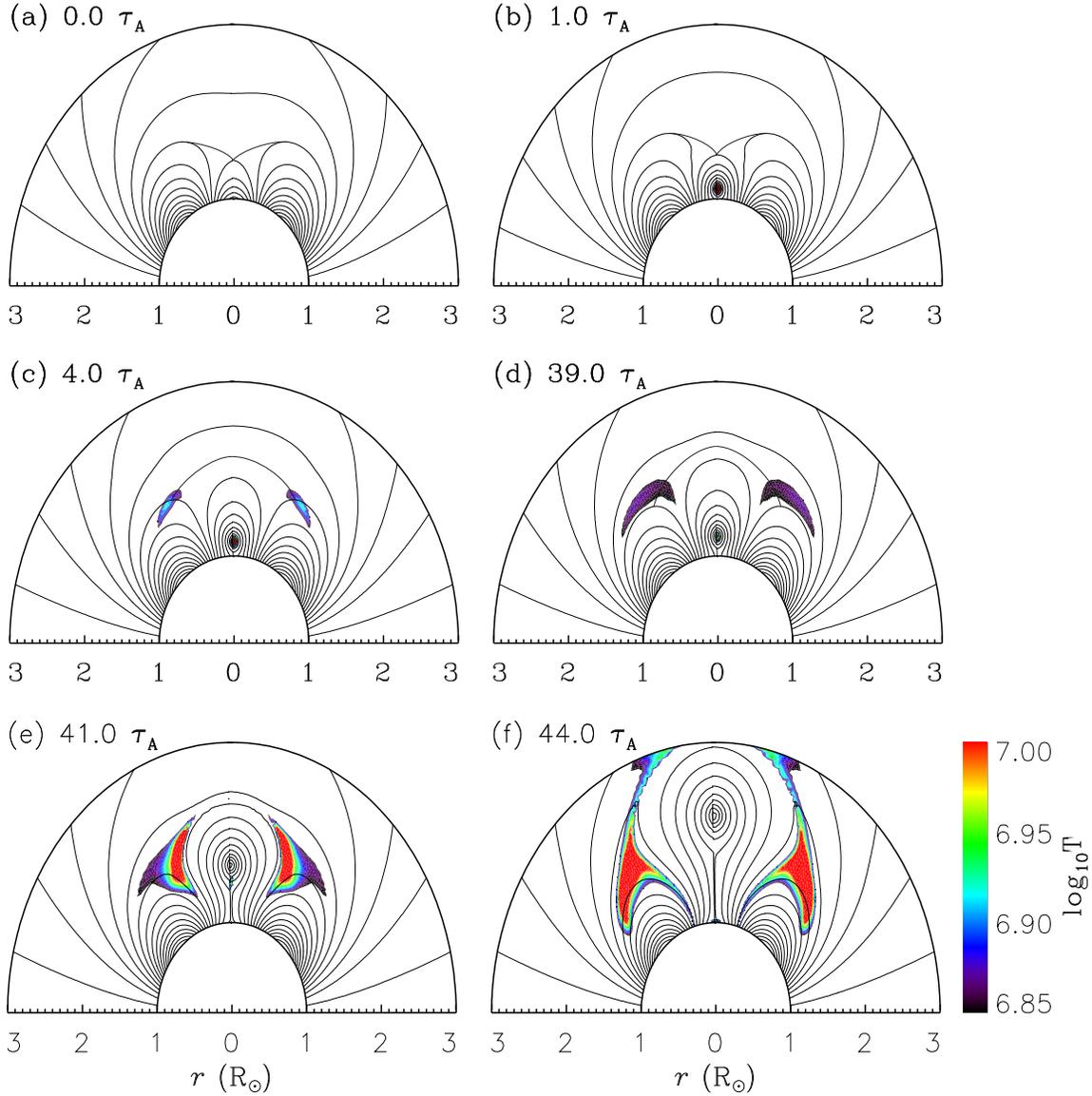}
  \caption{
  In order to show topology of the magnetic field and explosive
  flares clearly, magnetic configuration (black contours) and temperature distribution
  (color scale) are shown at six separate time, which are the initial
  quadrupole magnetic field (a), the flux rope just emerging into the solar corona (b),
  the start of the solar flares just after the first catastrophe (c), the end
  of the pre-flare just before the second catastrophe (d), magnetic reconnection
  just occurring in the vertical current sheet (e) and the end of the solar
  flares (f). The dark blue and bright red correspond to $7.08\times 10^{6}$
  and $1.0\times 10^{7}$K, respectively.
  }\label{Fig3}
\end{figure}

\subsection{Evolution of the solar flares}

Figure 4 shows that there are three different trends in the variation of current density
in the vertical current sheet. Accordingly, we believe that the solar flares usually consist
of three stages: pre-flare phase (4-39$\tau_{A}$), flare phase (39-41$\tau_{A}$) and post-flare
phase (41-44$\tau_{A}$). During the pre-flare phase, thermonuclear fusion has taken place
slowly and steadily inside the flux rope, for which $T_a$ has been much higher
than 7$\times10^{6}$ K for about 2.14 hours. Meanwhile, magnetic pinch process has maintained
the plasma temperature to keep the reaction going on. As a result, a large number of $^{2}H$
and $^{3}He$ accompanied by high energy gamma rays have been produced in the former two steps
of the p-p chain, which the duration of the pre-flare phase often determines production of
$^{3}He$.

\begin{figure}
  \centering
  \includegraphics[]{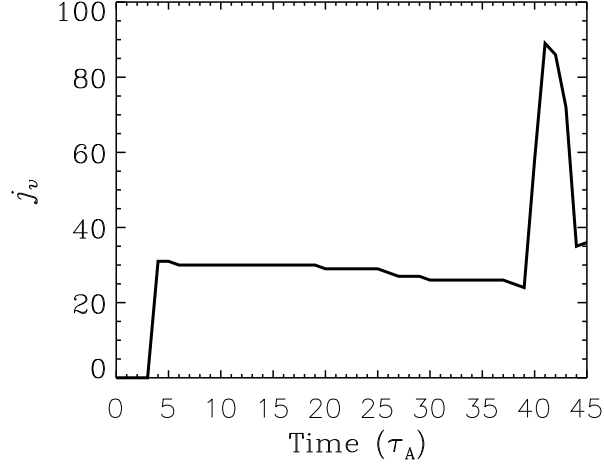}
  \caption{Current density of the vertical current sheet, $j_v$ vs. time, the unit
  of $j_v$ is $j_0$ = 1.65$\times 10^{-5}$A$\cdot$m$^{-2}$.
  }\label{Fig4}
\end{figure}

Subsequently, second catastrophe occurs just after 39$\tau_{A}$, at which the vertical
current sheet starts to stretch rapidly upwards, and then its current density increases
rapidly and gets the maximum value at 41$\tau_{A}$(Fig. 4). Due to magnetic flux rope upward
movement, plasma temperature inside the flux rope has been lowered, and the high temperature
zone has expanded at the footpoint on either side of the transverse current sheet (see Fig. 3d).
The violent nuclear explosion in the flux rope is triggered by the second catastrophe, and
magnetic reconnection occurs in the vertical current sheet while current density reaches its
maximum at 41$\tau_{A}$. In the meantime, magnetic reconnection in the transverse current sheet
produces two high temperature banding area (above 1$\times10^{7}$ K), for which a large number
of accelerated charged particles bombard the lower atmosphere along magnetic field lines
(see Fig. 3e). We further conclude that the flare phase begins with the second catastrophe and
ends with magnetic reconnection in the vertical current sheet.

While current density of the vertical current sheet reduces quickly and returns to
vicinity of the value before the second catastrophe, the transverse current sheet fully
reconnects and magnetic flux rope bursts out (CMEs) at 44$\tau_{A}$ (see Fig. 3f).
There are two large high temperature areas at both footpoints (greater than 1$\times10^{7}$ K), which are
the double hard X-ray sources. In the post-flare phase, CMEs are triggered by magnetic
reconnection in both vertical and transverse current sheets, and high temperature
products (such as high-energy $^{1}H$,$^{2}H$ and $^{3}He$) generated by the nuclear fusion
reaction inside the flux rope are ejected from corona into interplanetary space, simultaneously.
However, plasma macroscopic stability is a prerequisite to the solar flares during evolution
of magnetic flux rope system.

\subsection{Energy conversion in the solar flares}

Solar flares are high energy astrophysical events that occur frequently in the solar
atmosphere accompanied by high-energy protons and anomalous abundance of $^{3}He$ with
gamma rays. Our simulation result shows the details of the energy conversion during
the solar flare process given in Table1, which magnetic energy is converted into
thermal energy, and hydrogen burns to release nuclear energy, at last, some of the
remaining magnetic energy drives coronal mass ejections into space.

\begin{table}
\begin{center}
\caption[]{Calculated magnetic flux rope system parameters at different times.}\label{table:1}
\begin{tabular}{c c c c c c c c}
\hline\noalign{\smallskip}
{$\tau_{A}$} & {$T_a$} & {$\rho_a$} & {$n_a$} & {$R_{a}$} & {$E_{H}$} & {$E$} & {$E_r$} \\
{} & {($10^{6}$K)} & {($10^{-11}$$kg\cdot m^{-3}$)} & {($10^{16}$$m^{-3}$)} & {($10^{8}$m)}
& {($10^{-16}$J)} & {($E_0$=5.563$\times 10^{28}$J)} & {($E_0$)} \\
\hline\noalign{\smallskip}
0  & 3     & 1.67  & 1     & 0     & 1.036  & 1.476 & 0      \\
1  & 34.94 & 3.597 & 2.154 & 0.802 & 12.06  & 1.730 & 0.1854 \\
41 & 10.67 & 2.921 & 1.749 & 1.564 & 3.683  & 1.707 & 0.1143 \\
\noalign{\smallskip}\hline
\end{tabular}
\end{center}
\end{table}

As magnetic flux emerges into the solar corona, the magnetic flux rope system gains magnetic
energy. About 1.413$\times 10^{28}$J of the added magnetic energy is used to maintain
balance of the flux rope system, that is why the magnetic confinement structure remains
40$\tau_{A}$ in the corona. Besides, about 5.390$\times 10^{25}$J magnetic energy is converted
into thermal energy of the plasma inside the flux rope by an adiabatic compression process
at about 1$\tau_A$. Thus plasma inside the flux rope has been heated to 3.494$\times10^{7}$K,
which $^{1}H$ has obtained very high average kinetic energy, and then burns in the flux rope of
radius 0.802$\times10^{8}$m. We assume that is the core region of magnetic field heating
plasma in the flux rope.

\begin{figure}
  \centering
   \includegraphics[]{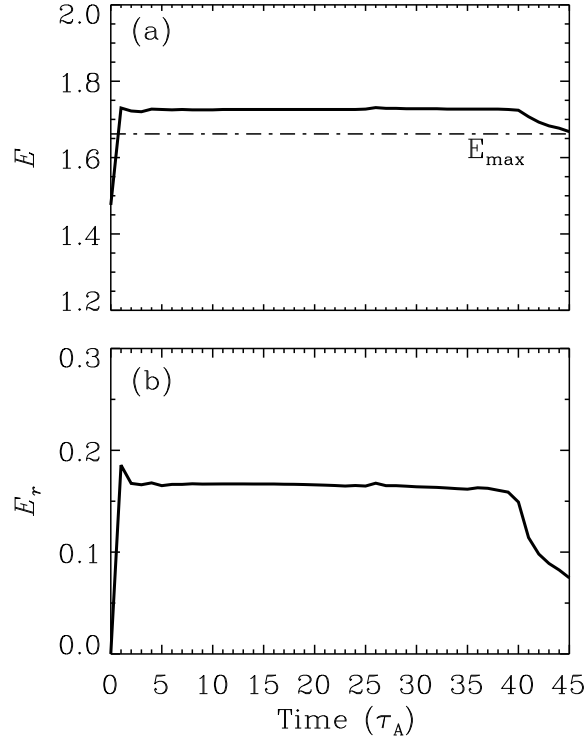}
   \caption{
   Magnetic energy of the system, E vs. time (a) , that of the flux rope,
  $E_r$ vs. time (b), and unit of the magnetic energy is
  $E_0$ = 5.563 $\times 10^{28}$ J.}
   \label{Fig5}
\end{figure}

For values of mean temperature $T_{f}$=1.625$\times10^{7}$K and mean density
$\rho_{f}$=1.950$\times10^{-11}$$kg\cdot m^{-3}$ in core of the flux rope during process of
thermonuclear fusion, we use the above equation (5) to calculate the rate of energy production in
the flux rope at the solar surface, $\epsilon_{f}$ = 3.46$\times10^{-4}$ erg $\cdot m^{-3}\cdot s^{-1}$.
Even with very low fusion rate, the amount of nuclear energy produced
also depends on both the timing of the fusion and the size of the fusion region ($m^{-3}$). Thus, from
1$\tau_{A}$ to 41$\tau_{A}$, the total amount of nuclear energy produced in the flux rope is estimated
at about 3.045$\times10^{-7}$J$\cdot m^{-3}$. Furthermore, we can estimate that about
2.746$\times10^{5}$($m^{-3}$) $^{3}He$ are produced during the solar flares. For different sizes of
nuclear reaction regions ($m^{3}$), we can calculate the nuclear energy produced and the amount of $^{3}He$.
This seems to suggest that the solar flares maybe a magnetic confinement fusion process, in which
thermonuclear reaction of hydrogen is enough to produce up to a factor of ten thousand of $^{3}He$ even
in the extremely low rate of energy production for several hours in the solar magnetic active region.
Obviously, the thermonuclear fusion produces not only $^{3}He$ but also high energy gamma rays that is
used to be a considerable puzzle observational phenomenon.

Figure 5 shows that magnetic energy of the flux rope system, E, and energy of the flux rope,
$E_r$, as a function of time, have the similar trend. A dot-dashed line in Figure 5a denotes the
corresponding partly open field energy $E_m$=1.662. It should be noted that the magnetic energy
remains stablely until 41$\tau_{A}$, which magnetic reconnection occurs in the vertical current
sheet and triggers the CMEs. While the transverse current sheet fully reconnects at about 44$\tau_{A}$,
magnetic energy of the system drops from 1.707 to 1.677. Meanwhile about 8.345$\times10^{26}$J of
magnetic energy in the system is released suddenly, and the magnetic flux rope is thrown out with
massive high energy particles (CMEs).

\section{Conclusions}
\label{sect:conclusion}

Our result suggests that the solar flares usually represented in three phases
(pre-flare, flare and post-flare) may be a process of magnetic confinement nuclear
fusion in solar atmosphere. After the first catastrophe, the
magnetic confinement structure appears in the form of a magnetic flux rope with
a vertical current sheet below and a transverse current sheet above. Between the
first and second catastrophe (pre-flare phase), the plasma heated by adiabatic
compression process in the flux rope has been involved in the steady thermonuclear
fusion. From the second catastrophe to magnetic reconnection just in the
vertical current sheet (flare phase), the solar flares manifest the process of the
transient nuclear explosion accompanied by high energy gamma rays. While magnetic
reconnection occurs in both the vertical and the transverse current sheets
(post-flare phase), thermonuclear fusion products (including $^{3}He$ and other
high energy particles) with magnetic flux rope and plasma in the corona are
injected into interplanetary space, simultaneously (CMEs).

We conclude that the solar flares are probably resulted from loss of equilibrium
in the complex magnetic configurations, and produced by magnetic confinement nuclear
fusion of the plasma during the ideal MHD process in solar atmosphere. Magnetic
reconnection is a key element of the solar flares, which occurs in an impulsive way
in the solar corona. Furthermore, magnetic reconnection may be a general phenomenon closely
coincided with the nuclear fusion explosion, and triggers the CMEs without plasma
macroscopic instability. In addition, the high energy particles involved in the solar
flare process maybe accelerate not by some kinds of acceleration mechanism, but by high
temperature.

\begin{acknowledgements}
The authors are greatly indebted to the anonymous referee for
helpful comments and valuable suggestions on the manuscript.
\end{acknowledgements}


\label{lastpage}

\end{document}